\documentclass{elsart}

\usepackage{amssymb}
\usepackage{epsf,colordvi,color,amsbsy}
\usepackage[dvips]{graphicx}
\begin{document}

\begin{frontmatter}

\title{On the correlation between air temperature and the core Earth processes: Further investigations using a continuous wavelet analysis}

\author[]{Stefano Sello \corauthref{}}

\corauth[]{stefano.sello@enel.it}

\address{Mathematical and Physical Models, Enel Research, Pisa - Italy}

\begin{abstract}
In a recent article by Dickey, J. O., Marcus, S.L. and O. de Viron, 2011, the authors show evidences for correlations in the multi-decadal variability of Earth's rotation rate [i.e., length of day (LOD)], the angular momentum of the core (CAM), and natural surface air temperature (SAT). Previous investigators have already found that the LOD fluctuations are largely attributed to core-mantle interactions and that the SAT appears strongly anti-correlated with the decadal LOD. As the above authors note, the cause of this common variability needs to be further investigated and studied. In fact, "since temperature cannot affect the CAM or LOD to a sufficient extent, the results favor either a direct effect of Earth's core-generated magnetic field (e.g., through the modulation of charged-particle fluxes, which may impact cloud formation) or a more indirect effect of some other core process on the climate-or yet another process that affects both". The main aim of the present research note is to further support the above results, using a proper continuous wavelet analysis, and to evidence some early detected time-period relations existing between LOD and SAT data. 
\end{abstract}
\end{frontmatter}

\section{Introduction}
We recall here some brief introductive considerations of the argument under study, as well reported in the reference article by Dickey, J. O., Marcus, S.L. and O. de Viron, 2011.  The length of day (i.e., the time needed by Earth to make one full rotation) fluctuates around its mean 24-h value, over a broad range of periods. Earth rotation is observed with a very high accuracy from precise geodetic techniques such as very-long-baseline interferometry, which allows for the determination of the length of day (LOD) at the 0.02-ms level (less than 1 cm at the equator). Earth rotation data collected before 1975 are the results of traditional optical observations; thus uncertainties during this period are significantly larger than for current determinations. The observed fluctuation of the LOD can be interpreted, at seasonal time scales, in terms of angular momentum exchange between the solid Earth and the superficial fluid layers (i.e. the atmosphere and the ocean), without the need to invoke influences from the core. For the atmosphere, variations in global angular momentum are dominated by the effect of the winds (the wind term), with the moment-of-inertia changes (the pressure term) being an order of magnitude smaller (Hide and Dickey 1991; Rosen 1993). The ocean contributions, being yet smaller than the pressure term, are significant in the closure of the angular momentum balance (Marcus et al. 1998). On the other hand, at decadal and longer time scales (multi-decadal), the main source of rotational variation is the interaction between the mantle and core, as substantiated by the significant correlation between the low-degree zonal components of the magnetic field and the LOD (Jault and Le Mouel 1991). The decadal LOD variations (e.g., $\sim 4 $ ms around 1900) are too large in amplitude to be explained by the atmosphere ($\sim 1 $ ms in amplitude).  The long-term global mean Earth surface air temperature (SAT) is significantly anti-correlated with decadal and longer LOD (e.g., Lambeck and Cazenave 1976). Variability at periods of 60-80 yr has been well established in LOD (e.g., Jault and Le Mouel 1991; Roberts et al. 2007) and in the core angular momentum (CAM), using both observational data (Zatman and Bloxham 1997; Dickey and de Viron 2009) and theoretical studies (Mound and Buffett 2007). 

It is well known that there are no direct observations of the fluid core motions. However, geomagnetic fields, generated by the dynamo inside the fluid core, can be used to infer the motion field and its velocity by utilizing several basic assumptions and the induction equation.  In fact, if the solid Earth-core system is assumed to be isolated, the changes of the solid-Earth angular momentum, and thus its rotation speed and the associated LOD, can be deduced from changes of the CAM. 
The main result of the above authors is that LOD, CAM and corrected (i.e. for anthropogenic effects) global mean air temperature show strong evidence of multi-decadal variability, with a period in the broad range: 60-80-yr. Oscillations in global temperatures with periods in the 65-70-yr range were originally reported by Schlesinger and Ramankutty (1994). Subsequent observational studies and simulations with coupled atmosphere-ocean models have found similar multi-decadal climatic modes, typically originating in the North Atlantic Ocean; however, the excitation source or sources of these oscillations have not been unambiguously identified (Knight 2009). The authors work suggests that the same core processes that are known to affect Earth's rotation and magnetic field (Roberts et al. 2007) may also contribute to the excitation of such modes, possibly through geomagnetic modulation on regional as well as global scales (Usoskin et al. 2008).  For more details see the full article by Dickey, J. O., Marcus, S.L. and O. de Viron, 2011.  

\section{Results}
Following the final comment of  the authors: "Further work remains to be done, especially in linking common modes of variability between Earth's subsystems and better describing the physical connections between them", here we used, as starting point, the curves of Fig. 1 in the original article (see the reported Figure 1), considering the -LOD behavior and the corrected temperature series from GISS beginning in 1880 and ending 2002.9.

\begin{figure}
\resizebox{\hsize}{!}{\includegraphics{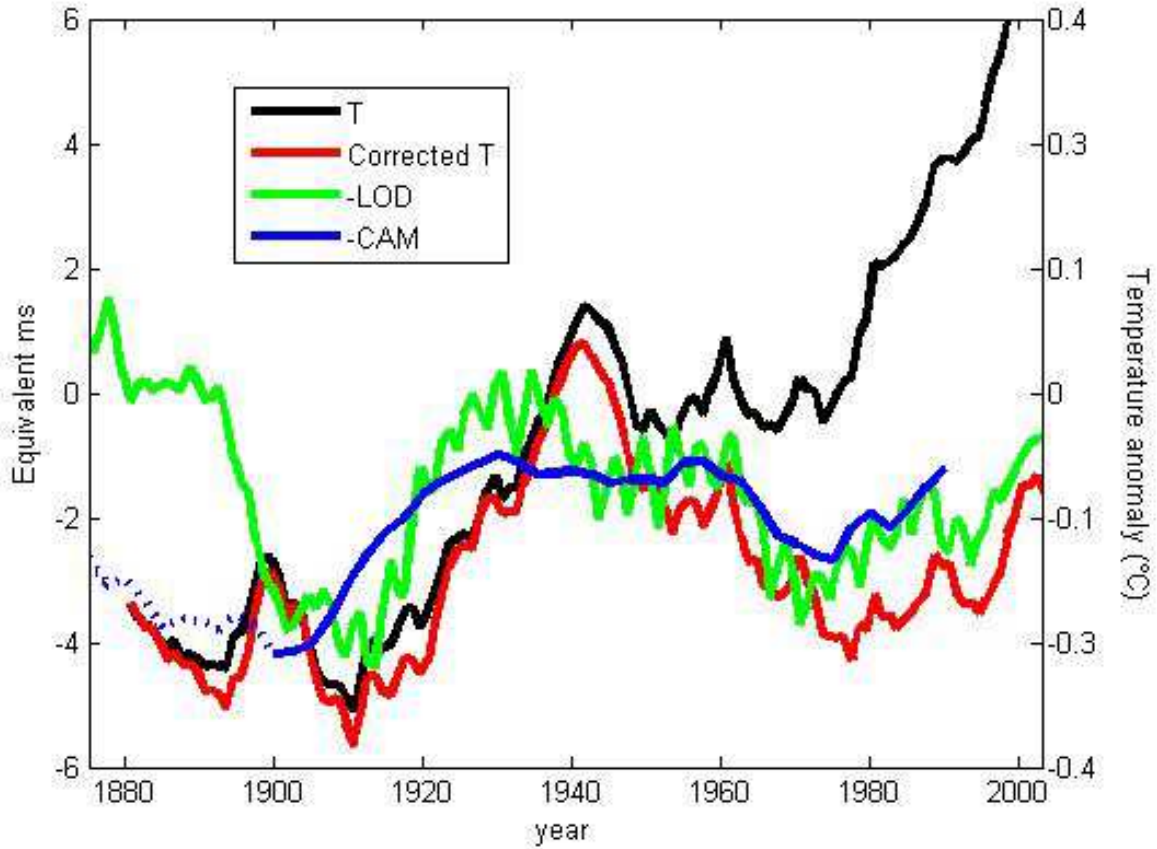}}
 \caption{The GISS observed and corrected temperature series (black and red lines, respectively) beginning in 1880 (cf. Hansen and Lebedeff, 1987). The LOD and CAM are given in units of equivalent milliseconds, with sign reversed so that positive anomalies correspond to increased rotational speed of the solid Earth, and the temperature series are given in units of 0.1 C. (From: Dickey, J. O., Marcus, S.L. and O. de Viron, 2011).}
 \label{fig1}
\end{figure}

The authors main results are:  "...the detection of a broadband variability centered at 78 yr (common variability ranges from 67 to 86 yr from SSA method). Oscillations in global temperatures with periods in the 65-70 yr are well known. Our work suggests that the same core processes that are known to affect Earth's rotation and magnetic field may also contribute to the excitation of such modes, possibly through geomagnetic modulation of near-Earth charged particle fluxes that may influence cloud nucleation processes, and hence the planetary albedo, on regional as well as global scales."

The main goal of the following wavelet analysis is to confirm the above global correlation between -LOD and Corrected Temperature (i.e. avoiding the anthropogenic component)  at multi-decadal scales and furthermore to better investigate time-frequency correlations between the series in order to strictly correlate the processes involved, observing their modes temporal behaviors. For mathematical details on this wavelet analysis, see: G. Ranucci and S. Sello, 2007.

The following Figure 2 shows the wavelet analysis performed on the first series: -LOD data
Note that the y-axis is shown in logarithmic scale in order to better see all the frequency range of interest involved (here selected from: 1 yr to: 121 yr). Black contour lines are boundaries of confidence regions at 95\%.

\begin{figure}
\resizebox{\hsize}{!}{\includegraphics{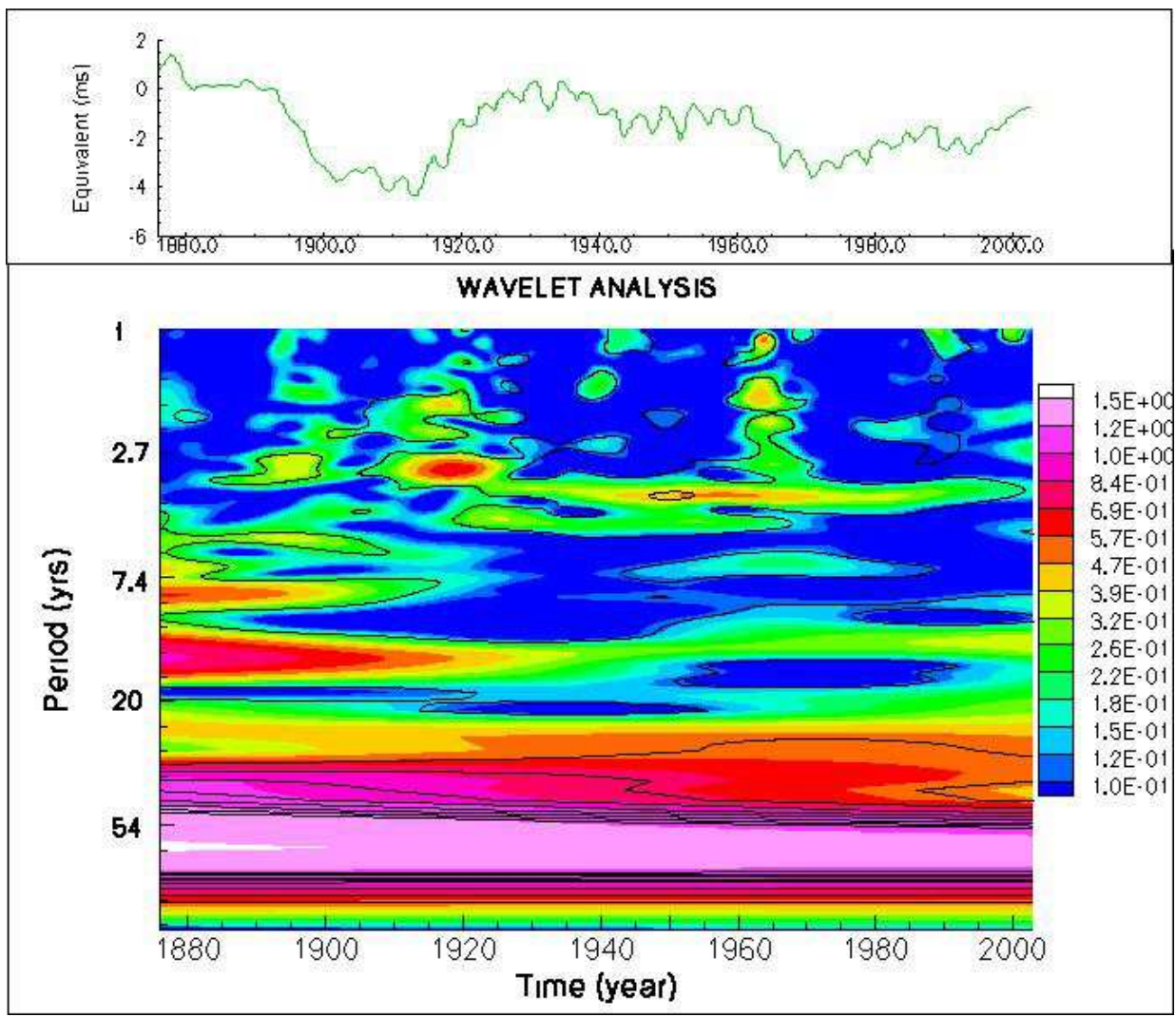}}
 \caption{Wavelet analysis of -LOD data}
 \label{fig2}
\end{figure}

Main results: we confirm a dominant and time-persistent frequency band in the periodicity range: 50-92 yrs, with a central peak at 65 yr, clearly visible up to 1900. Another persistent frequency band is near 34 yr. Further, other less persistent frequencies are: 33.8 - 13.8 - 8.41 - 3.7 - 3.0 and 1.7 yr. Some of these frequencies are well time localized and may be the object of further future investigations. 

The following Figure 3 shows the wavelet analysis performed on the second series: Corrected T data.
Note that the scales of the wavelet are the same as Figure 1 but the contour power levels here are rescaled to the different intensity energy level of the series. Black contour lines are boundaries of confidence regions at 95\%.

\begin{figure}
\resizebox{\hsize}{!}{\includegraphics{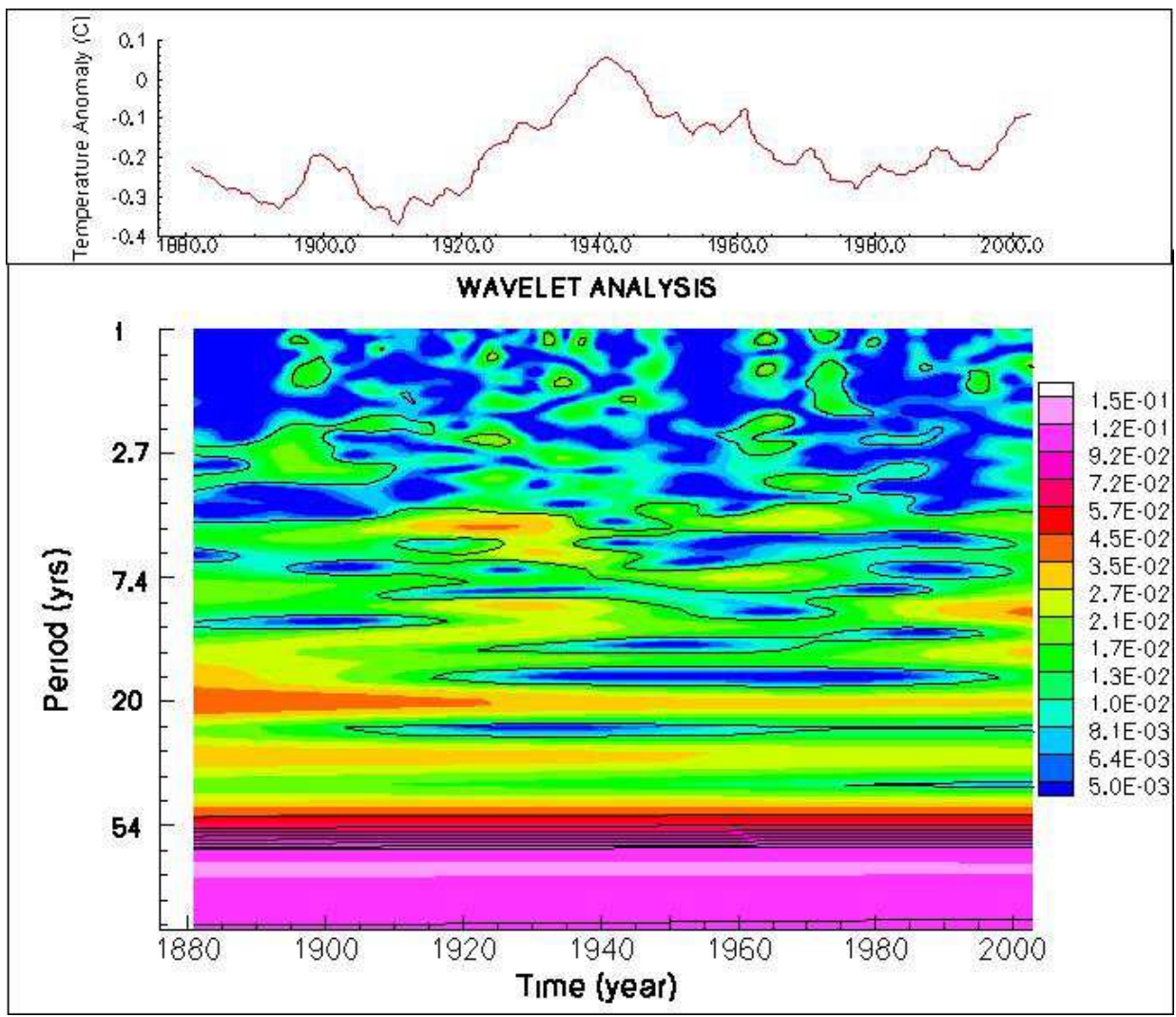}}
 \caption{Wavelet analysis of Corrected T data}
 \label{fig3}
\end{figure}

Main results: we confirm again a dominant and time-persistent frequency band in the range: 60-120 yrs, with a central peak at 78.25 yr, clearly visible in the full time range considered. Another persistent frequency band is near 30.8 yr. Further, other less persistent frequencies are now:  19.8 - 9.0 - 4.7 yr. Some of these frequencies are again well time localized but with interesting differences, both in relative intensity and/or time-frequency position from those of the -LOD case. These may be the object of further future investigations. A qualitative analysis of the main common periodicities detected both in -LOD and Temp series, leads to the preliminary conclusion that there is a general shift towards longer periodicities (from -LOD to Temp data) together with a phase shift of few years, probably related to the time response of atmosphere to changes of LOD. The excitation of similar modes shifted in time seems to exclude some external processes that affect Earth's core and climate simultaneously. In fact, the authors already note "an approximately 8-yr lag between the (negative) LOD and the corrected temperature; this lag agrees with the 8-yr lag between changes in Earth's rotational speed and surface geomagnetic field perturbations found by Roberts et al. (2007)".

The following Figure 4 shows the wavelet analysis performed on the third series: -CAM data
Note that the scales of the wavelet are the same as Figure 1. Black contour lines are boundaries of confidence regions at 95\%.

\begin{figure}
\resizebox{\hsize}{!}{\includegraphics{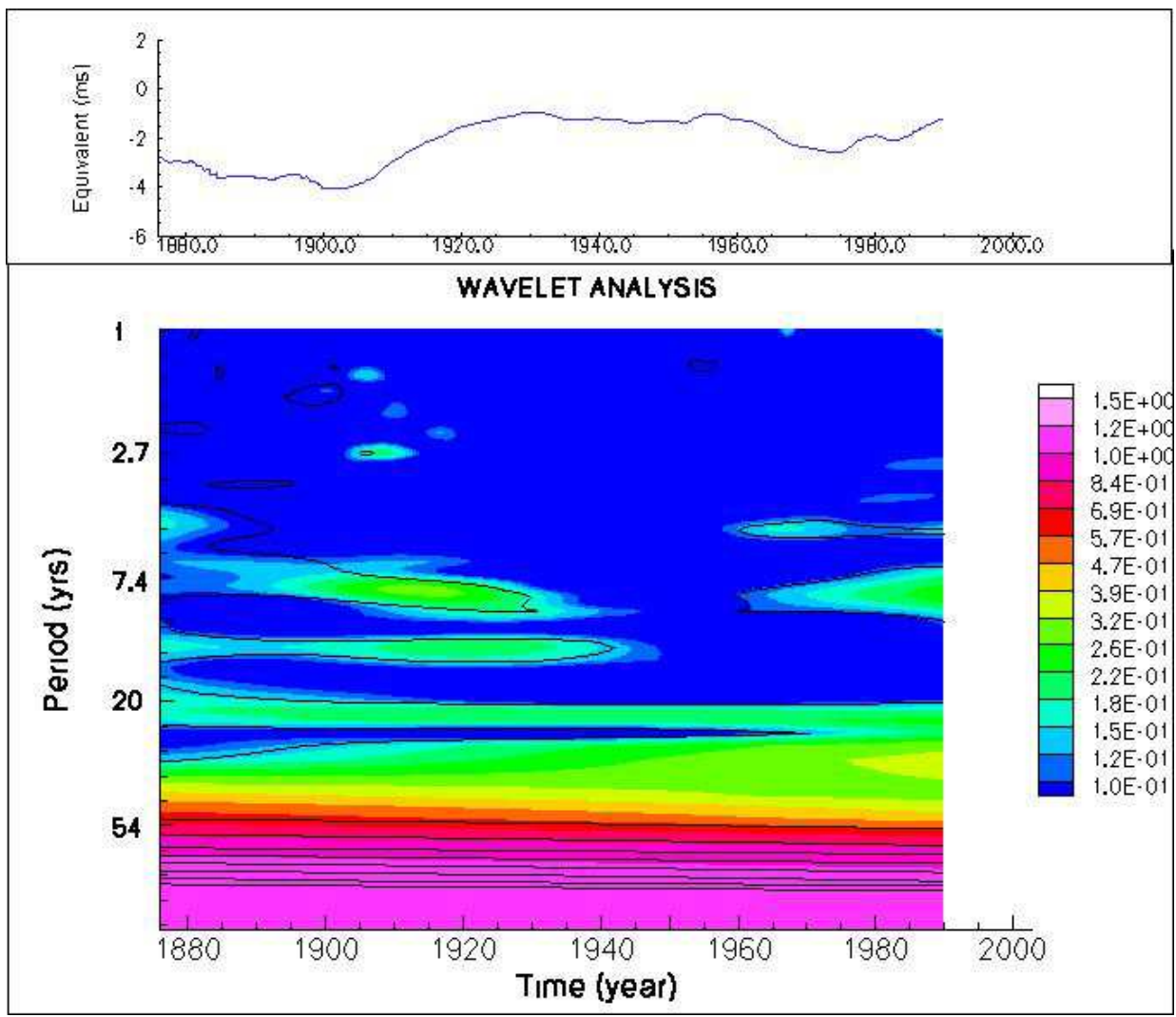}}
 \caption{Wavelet analysis of CAM data}
 \label{fig4}
\end{figure}

Main results: we confirm again a dominant and time-persistent frequency band in the long multi-decadal range. Here, due to the smoother and shorter CAM data, we were able to resolve, as maximum period, only the value: 55.7 yr, clearly visible in the full time range considered. Other significant periodicities are: 33.1 - 22.2 - 13.2 -  8.16 and 2.6 yr.  Some of these frequencies are again well time localized but with visible differences, mainly corresponding to different time positions from those of the -LOD case. However, there are clear common periodicities, as expected, with -LOD data. These localized periodicities may be the object of further future investigations.

\section{Conclusions}
The above wavelet analysis results support the idea, stated in the article by Jean O. Dickey, Steven L. Marcus and Oliver de Viron, that some sources of the (natural) global temperatures oscillations could be identified with the core processes that affect Earth's rotation and magnetic field. In particular, the excitation of similar modes shifted in time seems to exclude the possibility of external (e.g. solar) processes that affects the core and climate simultaneously. It seems more likely that some core processes, that induce mode excitations of rotation and magnetic fields, drive global air temperatures oscillations too. Through the multi-resolution wavelet analysis we can see that not only multi-decadal oscillations but also some high energetic decadal and sub-decadal well time-localized periodicities appear somewhat correlated and thus we need further investigations in order to give a correct interpretation. The time-frequency approach, available e.g. with a proper wavelet analysis, could help further detailed investigations and interpretations on this interesting suggested relation.

\section{References}
Dickey, J. O., Marcus, S.L. and O. de Viron:, 2011 Air Temperature and Anthropogenic Forcing: Insights from the Solid Eart, J. Climate, 24, 569-574.

Dickey, J. O., and O. de Viron, 2009: Leading modes of torsional oscillation within the earth's core. Geophys. Res. Lett., 36, L15302, doi:10.1029/2009GL038386.

Hansen, J., and S. Lebedeff, 1987: Global trends of measured surface air temperature. J. Geophys. Res., 92, 13 345-13 372. 

Hide, R., and J. O. Dickey, 1991: Earth's variable rotation. Science, 253, 629-637. 

Jault, D., and J. L. Le Mouel, 1991: Exchange of angular momentum between the core and the mantle. J. Geomag. Geoelectr., 43, 111-119. 

Lambeck, K., and A. Cazenave, 1976: Long-term variations in the length of day and climatic change. Geophys. J. Roy. Astron. Soc., 46, 555-573. 

Marcus, S. L., Y. Chao, J. O. Dickey, and P. Gegout, 1998: Detection and modeling of nontidal oceanic effects on Earth's rotation rate. Science, 281, 1656-1659. 
Mound, J., and B. Buffett, 2007: Viscosity of the earth's fluid core and torsional oscillations. J. Geophys. Res., 112, B05402, doi:10.1029/ 2006JB004426. 

Ranucci, G. and Sello, S., 2007: Search for periodicities in the experimental solar neutrino data: A wavelet approach, Phys. Rev. D, 75, 7.

Roberts, P. H., Z. J.Yu, and C. T. Russell, 2007: On the 60-year signal from the core. Geophys. Astrophys. Fluid Dyn., 101, 11-35. 

Rosen, R. D., 1993: The axial momentum balance of Earth and its fluid envelope. Surv. Geophys., 14, 1-29. 

Schlesinger,M. E., and N. Ramankutty, 1994: An oscillation in the global climate system of period 65-70 years. Nature, 367, 723-726. 

Knight, J. R., 2009: The Atlantic multidecadal oscillation inferred from the forced climate response in coupled general circulation models. J. Climate, 22, 1610-1625. 

Usoskin, I. G., M. Korte, and G. A. Kovaltsov, 2008: Role of centennial geomagnetic changes in local atmospheric ionization. Geophys. Res. Lett., 35, L05811, doi:10.1029/2007GL033040. 

Zatman, S., and J. Bloxham, 1997: Torsional oscillations and the magnetic field within the earth's core. Nature, 388, 760-763. 

\end{document}